\documentclass[preprint]{elsarticle}
\usepackage{epsfig}
\usepackage[english]{babel}
\usepackage{amssymb}

\begin{document}
\begin{frontmatter}
\title{A system of beam energy measurement based on the Compton backscattered
laser photons for the VEPP-2000 electron-positron collider}
\author[binp]{E.V.~Abakumova}
\author[binp,ngu]{M.N.~Achasov\corref{cor}}
\ead{achasov@inp.nsk.su}
\author[binp,ngu]{D.E.~Berkaev}
\author[binp]{V.V.~Kaminsky}
\author[binp,ngu]{I.A.~Koop}
\author[binp,ngu]{A.A.~Korol}
\author[binp]{S.V.~Koshuba}
\author[binp]{A.A.~Krasnov}
\author[binp,ngu]{N.Yu.~Muchnoi}
\author[binp,ngu]{E.A.~Perevedentsev}
\author[binp]{E.E.~Pyata}
\author[binp]{P.Yu.~Shatunov}
\author[binp]{Yu.M.~Shatunov}
\author[binp]{D.B.~Shwartz}

\cortext[cor]{Corresponding author}
\address[binp]{Budker Institute of Nuclear Physics, Siberian Branch of the
Russian Academy of Sciences, 11 Lavrentyev,
Novosibirsk 630090, Russia}
\address[ngu]{Novosibirsk State University,
Novosibirsk 630090, Russia}

\begin{abstract}
The beam energy measurement system for the VEPP-2000 electron-positron 
collider is described. The method of Compton backscattering of $CO$ laser
photons on the electron beam is used. The relative  systematic uncertainty of 
the beam energy determination is estimated as $6\cdot10^{-5}$. It was obtained 
through comparison of the results of the beam energy measurements using
the Compton backscattering and resonance depolarization methods. 
\end{abstract}

\end{frontmatter}

\section{Introduction}

\begin{figure}
\begin{center}
\includegraphics[scale=0.43]{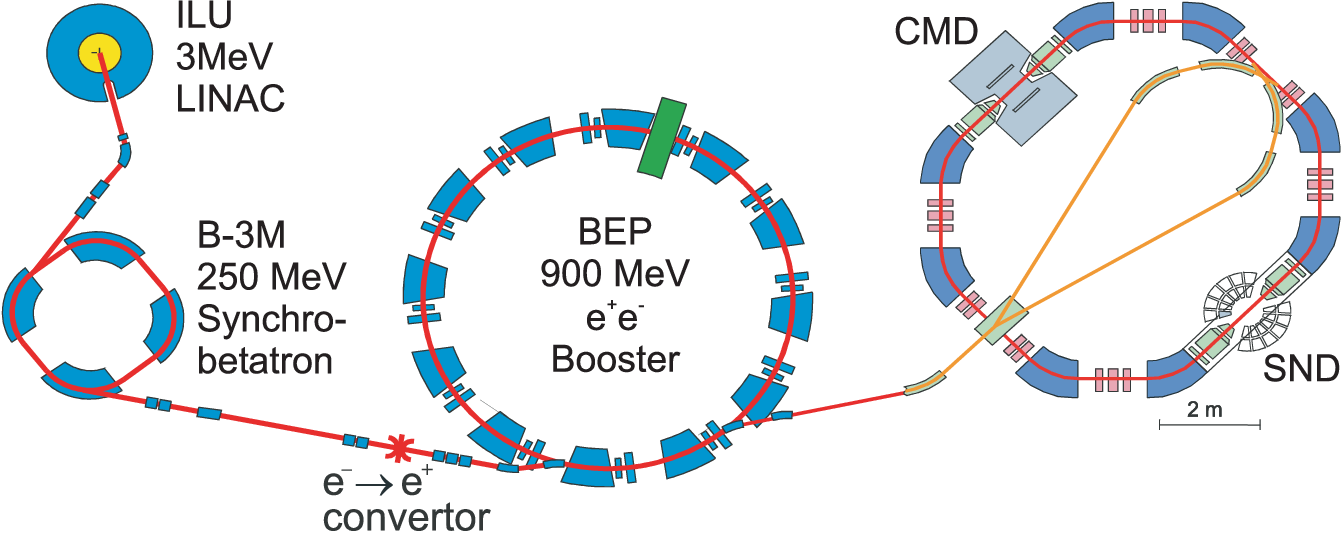}
\caption{VEPP-2000 accelerator complex.}
\label{vepp2000}
\end{center}
\end{figure}

The $e^+e^-$ collider VEPP-2000(BINP, Novosibirsk) \cite{vepp2000} for the
energy range $\sqrt{s}=0.4$ -- 2 GeV with a peak luminosity of
$10^{32}$cm$^{-2}$s$^{-1}$ was commissioned in 2007. The design of VEPP-2000
is based on the concept of round colliding beams, which allows to achieve
luminosity up to $10^{32}$cm$^{-2}$s$^{-1}$ in a bunch per bunch regime due to
suppression of the beam-beam tune shift \cite{rb}. The layout of the 
VEPP-2000 complex is shown in Fig.\ref{vepp2000}.

Currently the Cryogenic magnetic detector (CMD-3) and Spherical neutral 
detector (SND) collect data at VEPP-2000. The main goals of these experiments 
are
\begin{itemize}
\item
High precision measurements of the cross section of the processes of
$e^+e^-$ annihilation into hadrons.
These results are of paramount importance for the test of the  
Standard Model by a precise comparison of the experimental and 
theoretical predictions for the anomalous magnetic moment of the muon.
\item
Study of the light vector mesons $\rho,\omega,\phi$ and their excited
states $\rho^\prime, \omega^\prime, \phi^\prime, {\ldots} $
\item
Study of nucleon electromagnetic form factors near threshold in the process
$e^+e^-\to N\overline{N}$. 
\end{itemize}

The high accuracy of collider beam energy determination is crucial for a lot of
physical studies. For example, in order to measure the cross section of the
process $e^+e^-\to\pi^+\pi^-$  with accuracy better than 1\%, the beam energy
should be determined with a relative error of $10^{-4}$. The beam energy
measurement using Compton backscattering of monochromatic laser radiation on
the electron beam  (CBS method)  provides such
an accuracy, and allows to measure energy during 
data taking. This approach was developed and experimentally
proved in Refs.\cite{okp1,okp2,bessy1,bessy2,vepp4}.  At the BESSY-I storage 
ring the relative accuracy of energy measurement of about $10^{-4}$ 
was achieved for the beam energy of 800 MeV \cite{bessy1}. This accuracy 
was confirmed by comparison of the CBS measurement with the results obtained 
by the resonance depolarization (RD) method 
\cite{rd}. In collider experiments, the CBS method was applied at VEPP-4M 
\cite{vepp4} and the $\tau-charm$ factory BEPC-II \cite{bems}. 
A similar system was
proposed and constructed for VEPP-2000. In this paper, the system design and
performance are reported. The comparison of CBS and RD measurements 
has been performed.

\section{CBS method at VEPP-2000}

\begin{figure}
\begin{center}
\includegraphics[scale=0.7]{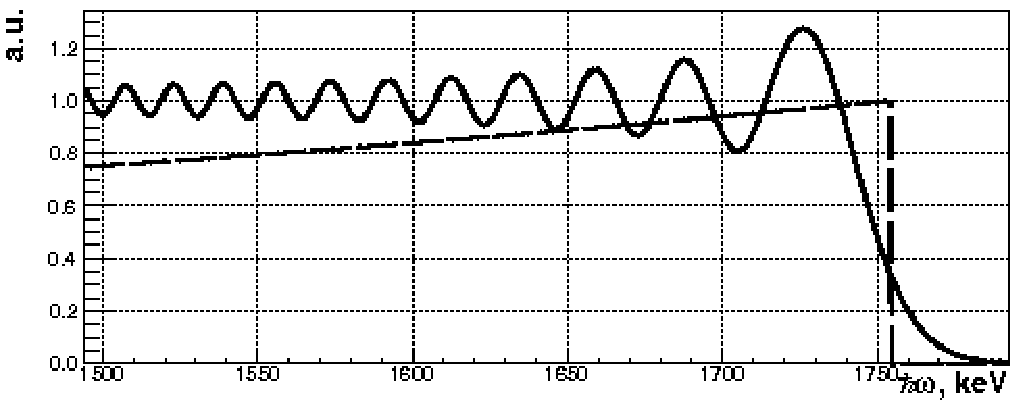}
\caption{The energy spectrum of scattered photons. The dashed line is 
the energy distribution according to the Klein-Nishina cross section with 
the aburpt edge given by scattering kinematics, the solid line is the energy 
distribution according to Eq.(\ref{ff}).}
\label{2d}
\end{center}
\end{figure}

In the previous works \cite{bessy1,bessy2,vepp4,bems} the CBS method 
was realized as follows. Laser light interacts with an electron beam at 
the straight
part of its  orbit, i.e. the angle $\alpha$ between initial particles is 
equal to $\pi$. The energy spectrum of back-scattered photons is described by
the Klein-Nishina cross section \cite{CN} with a sharp edge at the 
maximal energy
$\hbar\omega_{max}$ (Fig.\ref{2d}) due to kinematics of the Compton scattering.
Photons are detected by the High-Purity Germanium (HPGe) detector. The
ultra-high energy resolution ($\sim 10^{-3}$) of the HPGe detector allows a
relative statistical accuracy in the $\hbar\omega_{max}$ measurement 
to be at the level of $10^{-4}$ -- $10^{-5}$. The beam energy can be 
calculated using the measured $\hbar\omega_{max}$ value:
\begin{equation}
E=\frac{\hbar\omega_{max}}{2}
\biggl[1+\sqrt{1+\frac{m_e^2}{\hbar^2\omega_0\omega_{max}}}\biggr],
\end{equation}
where $\hbar\omega_0$ is the laser photon energy.

\begin{figure}
\begin{center}
\includegraphics[scale=0.6]{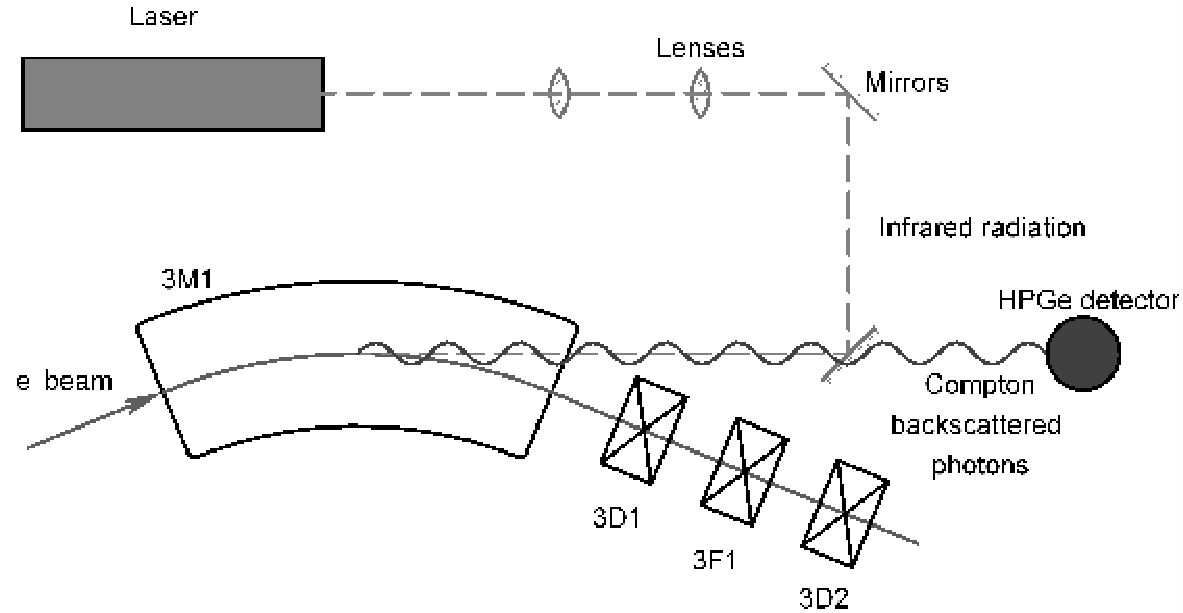}
\caption{Layout of the VEPP-2000 beam energy measurement system}
\label{layout}
\end{center}
\end{figure}

At VEPP-2000 the interaction of laser photons with electrons occurs 
inside the 3M1 
bending magnet at the curvilinear part of orbit. The layout of the beam
energy measurement system is shown in Fig.\ref{layout}. In this case the
energy spectrum of scattered photons (Fig.\ref{2d}) is described by the 
formulae 
\cite{PRL}:
\begin{equation}
\frac{\mathrm{d}\dot{N}_\gamma}{\mathrm{d}\hbar\omega}\propto
\nu\int\limits_{z}^{\infty}\mathrm{Ai}(z^\prime)\mathrm{d}z^\prime,
\label{ff}
\end{equation}
where
\begin{equation}
z=\biggl(\frac{u}{\chi}\biggr)^{2/3}\biggl(1-\frac{\kappa}{u}\biggr),
\end{equation}
\begin{equation}
\nu=\frac{1}{8}
\biggl\{ 2+\frac{u^2}{1+u}
-4\biggl[\frac{u}{\kappa}\biggr]^2 \biggr\},
\end{equation}
\begin{equation}
u=\frac{\hbar\omega}{(E-\hbar\omega)}, \mbox{~~}
\kappa=\frac{4E\hbar\omega_0}{m_e^2}, \mbox{~~}
\chi=\frac{EB}{m_eB_0}.
\end{equation}
Here $\mathrm{Ai}(z)$ is the Airy function, $\hbar\omega$ is the scattered
photon energy, $B$ is the dipole magnet field,
$B_0=m_e^2/\hbar c^2=4.414\times 10^9$ T. Taking into account the beam
energy spread, the energy distribution of the scattered photons (\ref{ff})
can be expressed as
\begin{equation}
\frac{\mathrm{d}\dot{N}_\gamma}{\mathrm{d}\hbar\omega}\propto
\mathcal{F}(\omega,E,B,\sigma_0)=e^{-\eta^6/24}
\int\limits_{z+\eta^4/4}^{\infty}e^{z^\prime\eta^2/2}
\mathrm{Ai}(z^\prime)\mathrm{d}z^\prime,
\label{fff}
\end{equation}
where
\begin{equation}
\eta\simeq\sigma_0\times{4\over 3}
\biggl(1+ {1\over 2} {\kappa\over u}\biggr)\biggl({u\over\chi}\biggr)^{2/3}.
\end{equation}
Here $\sigma_0$ is a relative beam energy spread. The beam energy $E$ is
obtained by a fit of the measured spectrum to the theoretical distribution
(\ref{fff}).

\section{The system of beam energy measurement for VEPP-2000}

The system of beam energy measurement  consists of the laser source, 
optical and laser-to-vacuum insertion systems to transport the laser beam 
into the interaction region where the laser beam collides with the electron  
beam, and the HPGe detector to measure backscattered photons.

\begin{figure}[t]
\begin{center}
\includegraphics[scale=0.40]{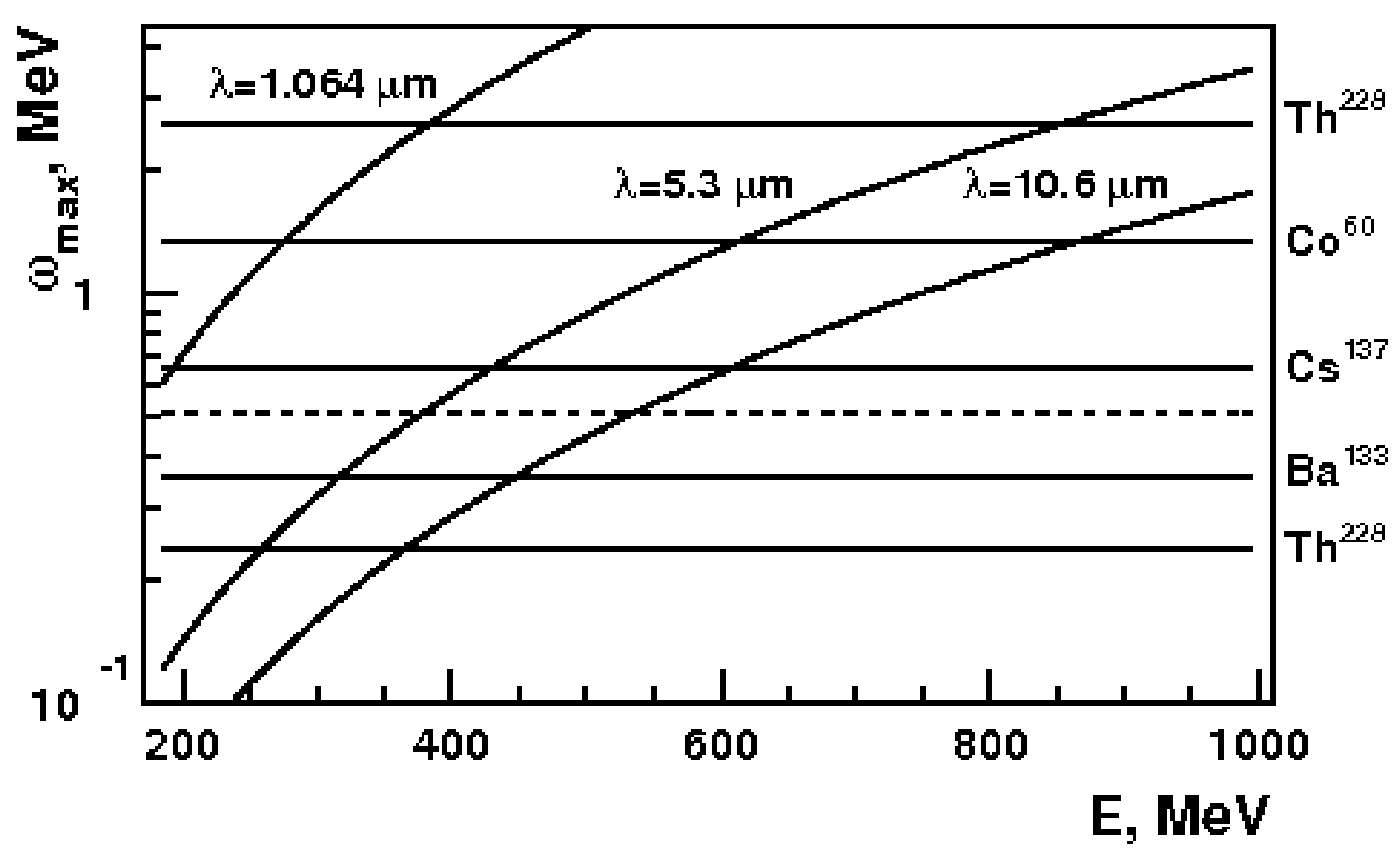}
\caption{Relation between $\omega_{max}$ and  $E$ for different
laser wavelengths. The solid lines are the energies of $\gamma$-active
radionuclide reference lines for the HPGe calibration. The dashed line shows
the photon energy 0.51 MeV from  $e^+e^-$ annihilation}
\label{ucmo}
\end{center}
\end{figure}

The source of initial photons is a PL3 $CO$ laser from Edinburgh Instruments.
The laser wavelength is calibrated with a relative accuracy of $10^{-3}$ by
a manufacturer, but this precision in not sufficient to measure beam energy 
with an accuracy of  $10^{-4}$. Therefore, a laser was additionaly calibrated 
with the  help of a high precision wavelength meter WS6-200 IR-III from High 
Finesse/$\mathring{A}$ngstrom. It was
found that the laser has a maximal power of 2W at the wavelength 
$\lambda_0=5.426468\pm 0.000005$ $\mu$m. This value agrees well with the table
data 5.426463 $\mu$m  for the wavelength of the P23 transition in 
a $CO$ molecule. 
Thus, the contribution of the laser wavelength accuracy to the 
relative error of the
beam energy determination is less than $10^{-6}$. The $CO$ laser is used
because in the entire VEPP-2000 energy range $E\simeq 0.2$ -- 1.0 GeV 
the energy of back scattered photons can be measured by the HPGe detector
($\hbar\omega_{max}\simeq 0.1$ -- 3.5 MeV). There exist $\gamma$-active 
radionuclides suitable  for a detector calibration in this energy range.
(Fig.\ref{ucmo}).

The total distance from the laser output aperture to the interaction region
is 773 cm. Two ZnSe lenses with focal lengths of $f_1=27$ cm and $f_2=43$ cm
focus a laser beam in such a way that its transverse  size at the interaction
region is 1 mm. The laser beam passes the lenses and is reflected by the 
mirror to a viewport in a vacuum pipe extension of the beam pipe. The 
mirror is installed on a special support that allows precise vertical and 
horizontal angular alignment by using stepping motors (one step equals 
$1.5\times 10^{-6}$ rad).

\begin{figure}[t]
\begin{center}
\includegraphics[scale=0.08]{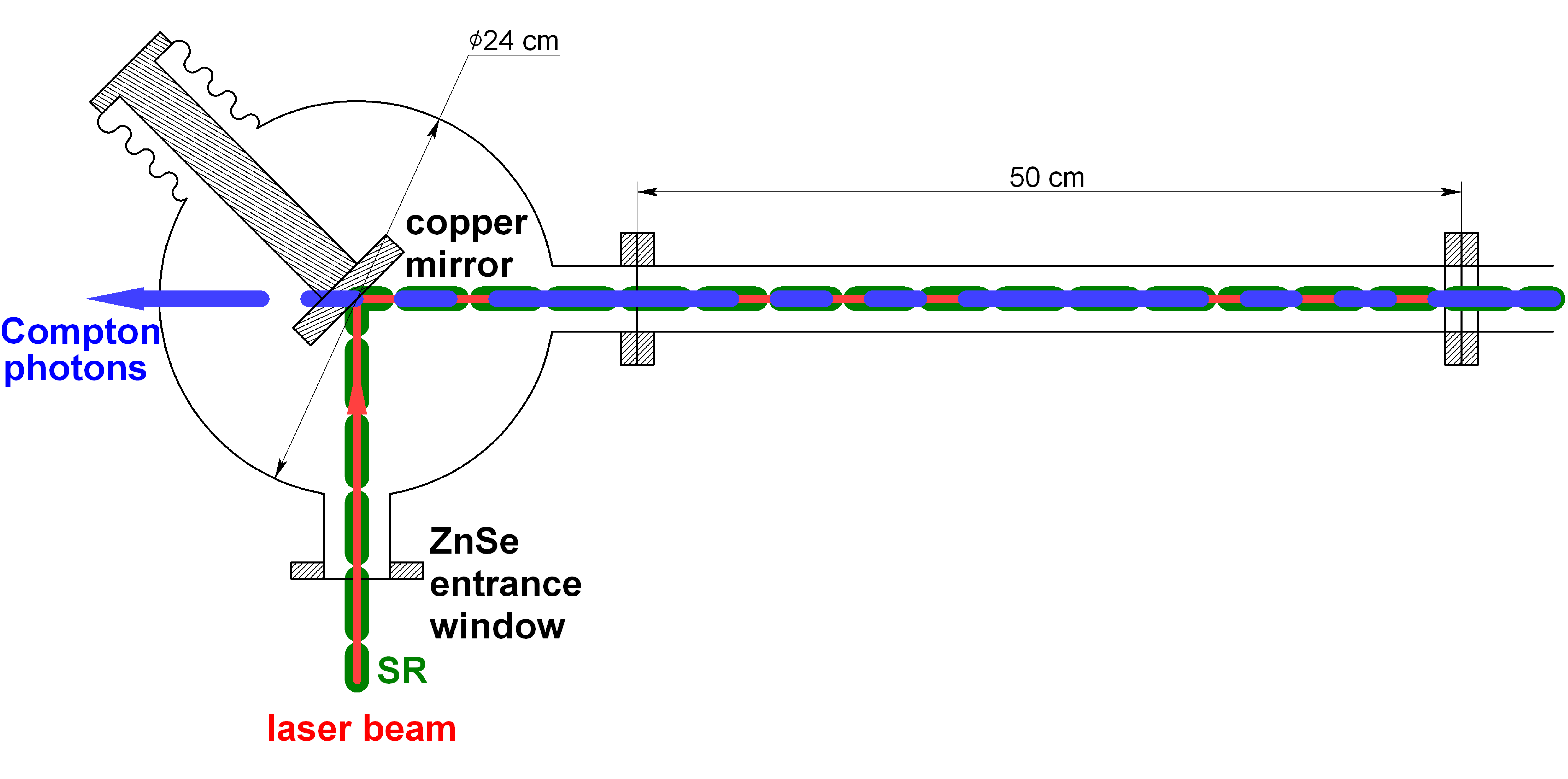}
\caption{Simplified schematic view of the laser-to-vacuum insertion system}
\label{laser-ins}
\includegraphics[scale=0.3]{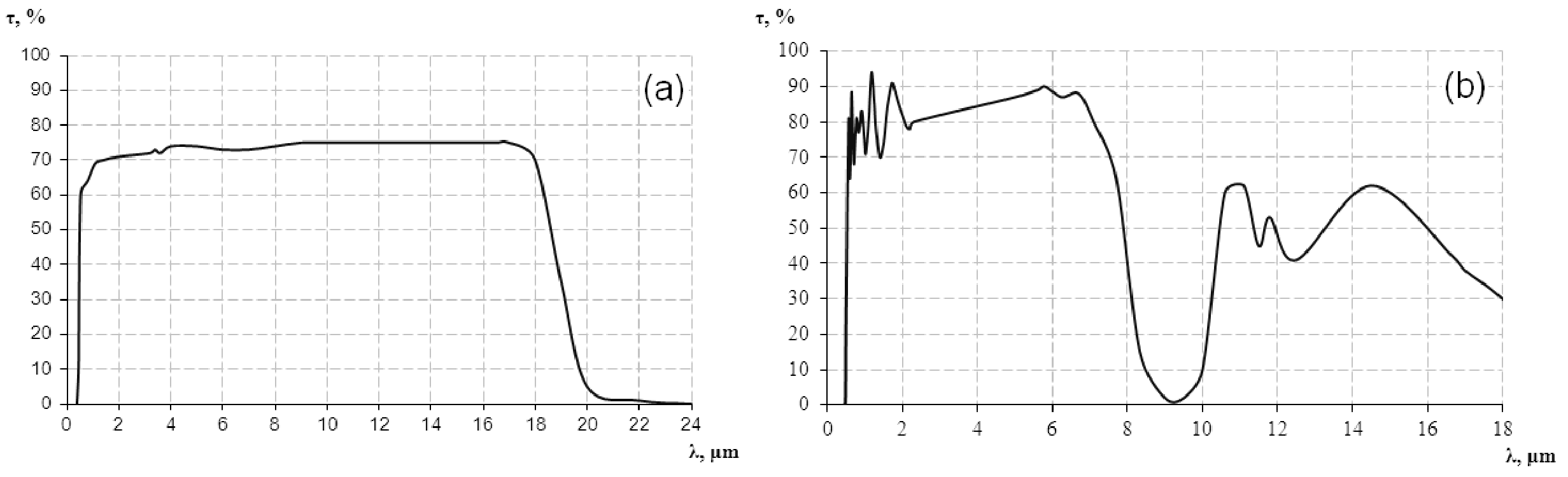}
\caption{The transmission spectra of ZnSe are shown for the a) 3 mm thick 
original plate; b) final product}
\label{znse}
\end{center}
\end{figure}

The insertion of the laser beam into the vacuum chamber is performed using the
laser-to-vacuum insertion system (Fig.\ref{laser-ins}). The system is a
special stainless steel  vacuum chamber with a ZnSe entrance viewport and 
copper mirror. The laser beam passes to the vacuum chamber through the entrance 
viewport and is reflected by an angle of $90^\circ$ at the copper 
mirror. After backscattering, the photons return to the mirror, pass through
it, leave the vacuum chamber, and are detected by the HPGe detector.

The design of the copper mirror was presented in Ref.\cite{bems,vac}. 
The viewport based on the ZnSe polycrystal provides:
\begin{enumerate}
\item
transmission spectrum from 0.45 up to 20 $\mu$m,
\item
baking out of the vacuum system up to 250$^\circ$C,
\item
very high vacuum.
\end{enumerate}

The viewport was  manufactured using a ZnSe crystal plate with a diameter of 
50.8 mm and thickness of 8 mm. The transmission spectrum of the plate is  shown
in Fig.~\ref{znse}(a). The ZnSe viewport design is similar to the design of 
the $GaAs$  viewport described in Refs.\cite{bems,vac,vp}. The transmission 
spectrum of the product is shown in Fig.~\ref{znse}(b).

After installation of the vacuum chamber at VEPP-2000 and pumping out, a 
pressure of  $3\times 10^{-10}$ Torr was obtained.

The optical elements of the system are adjusted using the synchrotron radiation
(SR) light of the electron beam. The copper mirror of the vacuum chamber 
and the 
mirrors of the optical system are adjusted in such a way that the SR light 
comes to the laser output window. Application of the ZnSe viewport in contrast
to GaAs one\cite{bems} makes system adjusting more convenient because  ZnSe 
is transparent for the visible part of SR.

At VEPP-2000 the n-type coaxial HPGe detector manufactured by ORTEC 
(model GMX25-70-A) is used for the system of energy measurement. 
It has a diameter of 
51.1 mm and  height of 72.9 mm. The energy resolution for the 1.33 MeV line of 
$^{60}Co$ is 1.9 keV (FWHM). The detector is connected to the multichannel 
analyzer (MCA) ORTEC DSpec Pro, which transfers data using the USB port of the
computer.

The HPGe spectrum has $2^{14}=16384$ channels. The bin error
for each channel is determined as
\begin{equation}
\Delta N =\sqrt{N+(\zeta N)^2},
\end{equation}
where $N$ is the number of counts in the channel and $\zeta$ corresponds
to the MCA differential nonlinearity, which is 0.02 according
to the MCA specifications.

In order to protect the HPGe detector from background, it is surrounded by
a 6 cm lead collimator with a 4 cm diameter hole in the direction of the beam 
scattered $\gamma $-quanta. The detector is arranged  in the collider's orbit
plane, at the distance of  225 cm from the interaction region of laser
radiation and an electron  beam.

The data acquisition system is similar to those described in Ref.\cite{bems}.
During data taking, the mirror is adjusted automatically to provide
a maximal photon/electron interaction efficiency using the feedback from the 
detector counting rate.

The data for the HPGe detector calibration -- peaks of the $\gamma$ sources and
peaks of the precise calibration pulse generator BNC model BP-5 with 
integrated nonlinearity $\pm 15$ ppm and jitter $\pm 10$ ppm are accumulated 
simultaneously with scattered photons. Generator signals are put to the
preamplifier with 12 different amplitudes covering the range of MCA and 
frequency of 1 Hz. The pulse shape is set in such a way that it is similar to 
the shape of the signal from a $\gamma$-quantum.

The system was operating in the test mode in April-December of 2012. In 2013
it started permanent operation in experiments at VEPP-2000.

\section{Data processing}

\begin{figure}
\begin{center}
\includegraphics[scale=2.1]{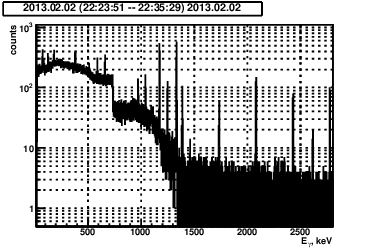}
\caption{The energy spectrum detected by the HPGe detector.  Several peaks,
corresponding to  the calibration generator, monochromatic $\gamma$-radiation 
radiative sources, and  the  edge of the Compton photon spectrum
 slightly below 750~keV are clearly seen.}
\label{cnekmp-1}
\includegraphics[scale=0.9]{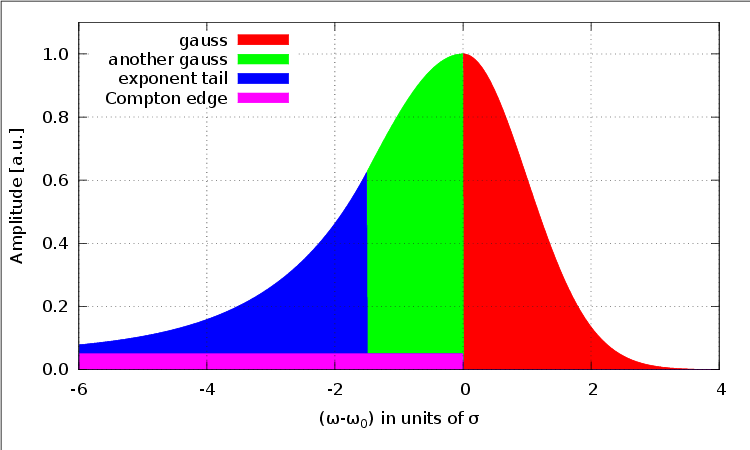}
\caption{HPGe detector response function}
\label{cnekmp-2}
\end{center}
\end{figure}

The processing of the spectrum (Fig.~\ref{cnekmp-1}) includes
calibration of the energy scale, Compton edge fitting and
determination of the beam energy. The procedure is similar to those described
in Ref.\cite{bems}.

\begin{table}
\caption{Radiative sources of $\gamma$-quantÁ which are used for the HPGe
detector calibration}
\begin{center}
\begin{tabular}{lc} \hline
Source & $E_\gamma$, keV \\
$^{208}Tl$     &  $583.191\pm0.002$ \\
$^{137}Cs$     &  $661.657\pm0.003$ \\
$^{60}Co$      & $1173.237\pm0.004$ \\
$^{60}Co$      & $1332.501\pm0.005$ \\
$^{208}Tl$     & $2614.553\pm0.013$ \\
\hline
\end{tabular}
\end{center}
\label{tab1}
\end{table}

The radiative sources used in this work for the HPGe detector scale 
calibration are  presented in Table~\ref{tab1}.
The goal of the HPGe detector calibration is to obtain the coefficients
needed for conversion of the MCA counts of the HPGe detector into the
corresponding energy deposition, measured in units of keV, as well as
to determine the parameters of the detector response function. The
following response function is used (Fig.\ref{cnekmp-2}):
\begin{equation}
f(x,x_0) = M \cdot
\left\{\begin{array}{ll}
\exp\biggl\{ {-{(x-x_0)^2\over 2\sigma^2}} \biggr\},
& 0< x+x_0<+\infty, \\
C+(1-C)\exp\biggl\{ {{(x-x_0)^2}\over{ 2(K_0\sigma)^2}} \biggr\},
& -K_0K_1\sigma < x-x_0\leq 0,  \\
C+(1-C)\exp\biggl\{ K_1\biggl({{x-x_0}\over{ K_0\sigma}}+{K_1\over 2}
\biggr) \biggr\},
& -\infty < x-x_0\leq -K_0K_1\sigma,  \\
\end{array}\right.
\label{m-gauss-norm}
\end{equation}
where $M$ is normalization, $x_0$ is the position of the maximum, $\sigma$ and
$K_0\sigma$ are RMS of the Gaussian distribution to the right and to the left 
of the $x_0$, respectively, $C$ is responsible for the small-angle Compton 
scattering of
$\gamma$-quanta in the passive material between the source and the detector,
$K_1$ is an asymmetry parameter.

\begin{figure}[t]
\begin{center}
\includegraphics[scale=0.6]{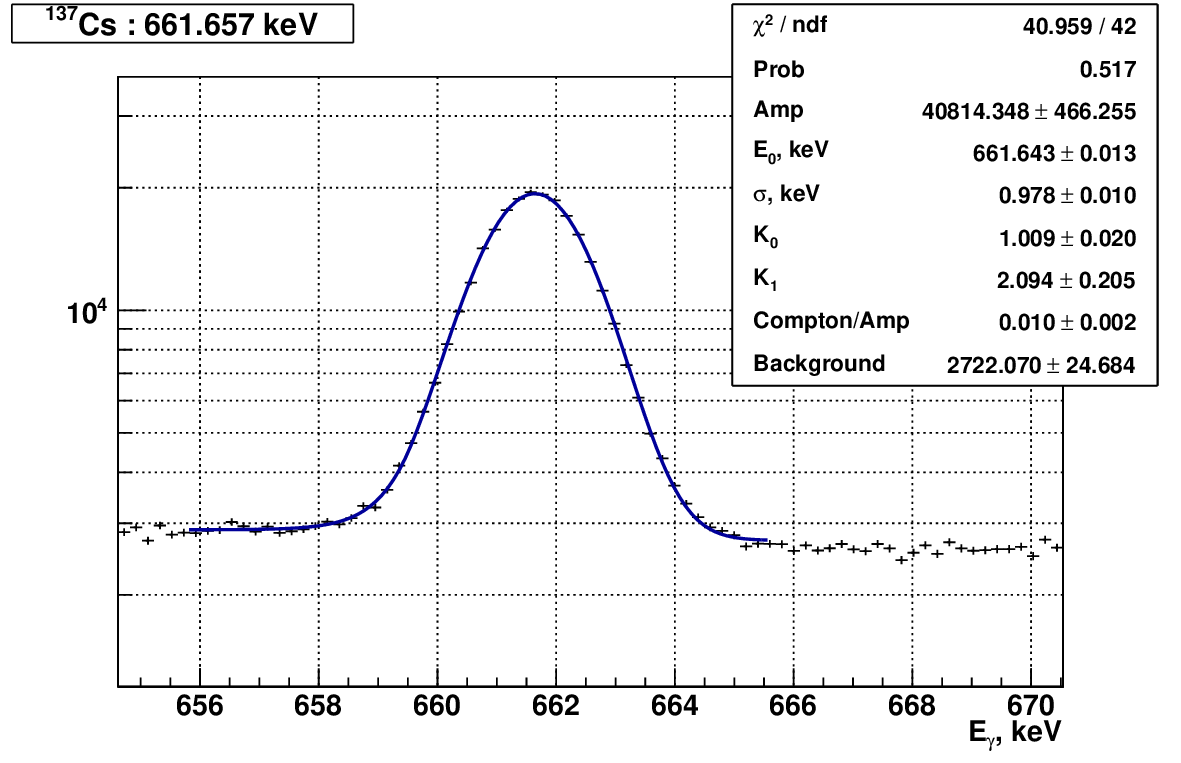}
\caption{The fit to the $^{137}Cs$ 661 keV peak}
\label{cnekmp-3}
\end{center}
\end{figure}

\begin{figure}
\begin{center}
\includegraphics[scale=2.1]{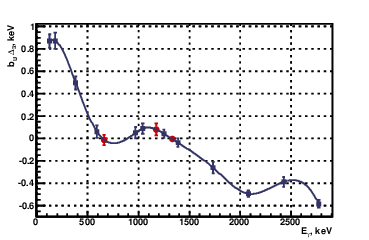}
\caption{Energy dependence of MCA scale nonlinearity. Squares -- generator
peaks, circles -- isotope peaks, curve - spline approximation.}
\label{linearity}
\includegraphics[scale=2.1]{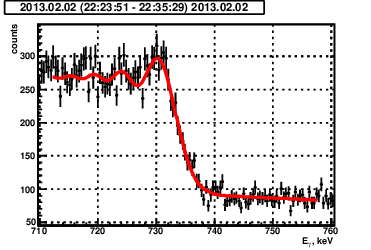}
\caption{The fit to the edge of backscattered photons spectrum, the beam
energy $E=458.508\pm 0.077$ MeV, $P(\chi^2)=0.45$.}
\label{cnekmp-4}
\end{center}
\end{figure}

The calibration procedure is as follows:
\begin{enumerate}
\item
Peak search and identification of the calibration lines (Table~\ref{tab1}).
\item
The peaks which correspond to calibration lines are fitted by a sum of
the signal and background distributions $f(x)+p$ (Fig.\ref{cnekmp-3}). The free
parameters of the fit are $x_0$, $\sigma$, $K_0$, $K_1$, $C$ and constant
coefficient $p$, which takes into account background. The generator peaks 
are well fitted by
the Gaussian distribution with a mean value $x_0$ and RMS  $\sigma$.
\item
Using generator data the nonlinearity of MCA scale is obtained:
\begin{equation}
U = a_N+b_NN+\Delta_3(N),
\label{pulsgen}
\end{equation}
where $U$ is a generator amplitude in V, $N$ is a corresponding amplitude 
in MCA counts, $a_N$ and $b_N$ are the linear coefficients, 
$\Delta_3(N)$ is a cubic spline, which takes into account ADC nonlinearity. 
Using the results of the
isotope peak approximation and values of  $U$ calculated from 
Eq.(\ref{pulsgen}) the coefficients $a_U$ and $b_U$ to convert the amplitude 
of the generator in the corresponding energy of $\gamma $-quanta in keV are
obtained:
\begin{equation}
\varepsilon = a_U+b_UU.
\label{genmev}
\end{equation}
Using (\ref{pulsgen}) and (\ref{genmev}), the formula for conversion of MCA
counts to the energy measured by the HPGe detector is:
\begin{equation}
\varepsilon=a_U+a_Nb_U+b_Ub_NN+b_U\Delta_3(N).
\end{equation}
The dependence of MCA nonlinearity on photon energy is shown in 
Fig.\ref{linearity}.
\item
Using the results of the isotope peak approximation, the energy dependence of
the response function parameters $\sigma$, $K_0$, $K_1$ and $C$ is determined.
\end{enumerate}

The edge of the backscattered photon spectrum (Fig.\ref{cnekmp-4}) is fitted
by the function:
\begin{equation}
S_2(\varepsilon,E,B,\sigma_0) = \int\limits_{\varepsilon}^{+\infty}
S_1(y,E,B,\sigma_0)\;dy\; + \mathcal{B}(\varepsilon),
\end{equation}
where the function
\begin{equation}
\mathcal{B}(\varepsilon)=p_0+p_1(\varepsilon-\hbar\omega_{max})
\end{equation}
approximates background and
\begin{equation}
S_1(\varepsilon,E,B,\sigma_0) = \int\limits_{-\infty}^{+\infty}
\mathcal{F}(\omega,E,B,\sigma_0)f(\varepsilon,\hbar\omega)\;d\hbar\omega\;.
\end{equation}
$\mathcal{F}(\omega,E,B,\sigma_0)$ is defined in Eq.(\ref{fff}).

\section{Comparison of CBS measurements with the RD method}

The beam energy determination using RD \cite{rd} is based on the relation 
between the electron energy and frequency $\Omega$ of its spin precession 
during the motion of the particle in the transverse magnetic field with 
a revolution frequency 
$\omega_s$:
\begin{equation}
E=\biggl(\frac{\Omega}{\omega_s}-1\biggr)\frac{\mu_0}{\mu^\prime}m_ec^2,
\label{depo}
\end{equation}
where $\mu_0/\mu^\prime$ is the ratio of the anomalous and normal parts of the
electron magnetic moment known with a relative accuracy of $2\times
10^{-10}$ \cite{pdg}. The frequency $\Omega$ can be obtained through resonant
depolarization of the polarized beam due to impact by an external
electromagnetic field with a frequency  $\omega_d$ such that
\begin{equation}
\omega_d\pm k\omega_s = \Omega (k\in\mathbb{Z}).
\label{freq}
\end{equation}
The relation (\ref{depo}) is violated in the presence of longitudinal
magnetic fields. Therefore, to measure the energy of the VEPP-2000 by RD the
collider focusing solenoids and solenoid of CMD-3 detector are swiched off.
Measurements are carried out at two energy points close to 458 and 509 MeV.

The positron beam with the energy of 800 MeV is polarized due to the 
Sokolov-Ternov effect (radiation polarization) \cite{baier} in the booster of 
electrons and positrons (BEP, Fig.\ref{vepp2000}). The time of polarization is 
about 50 minutes. During polarization the beam circulates in the booster for 2
hours. After that the beam energy is lowered to the value close to 485 or 509
MeV and the beam is injected to VEPP-2000. The positron beam 
is used for RD measurements, because positrons  both in  the booster
and collider are circulating in the same direction. Therefore, after injection
to VEPP-2000 the direction of the positron spin  does not change due to 
radiation polarization, in contrast to electrons. Since the electron and 
positron beams 
circulate in the same vacuum chamber in the same magnetic field and 
in the absence of transverse electric fields, the results of depolarization 
for them are identical  \cite{g-2e}.

The destruction of polarization is provided by a high-frequency depolarizer 
based on striplines, which is installed in the injection section of VEPP-2000.
In order to detect the moment of depolarization the process of intrabeam
scattering (Touschek effect) is used. The cross section of the scattering of
polarized electrons is smaller than for unpolarized ones. Therefore, after
depolarization the number of particles scattered out of the beam rises.
The scattered electrons are detected using the coincidence signal of two
scintillation counters, installed in the  internal and external  parts of
the VEPP-2000 straight section. While scanning the depolarizer frequency, an
approximately 2.5 \% rise of the counter counting rate is observed at the
moment of polarization destruction. From the time of the counting rate change
the value of the frequency $\omega_d$ is obtained and then, using
Eq.(\ref{depo}) and Eq.(\ref{freq}), the beam energy $E_ {RDP}$ is calculated.

\begin{figure}[t]
\begin{center}
\includegraphics[scale=0.5]{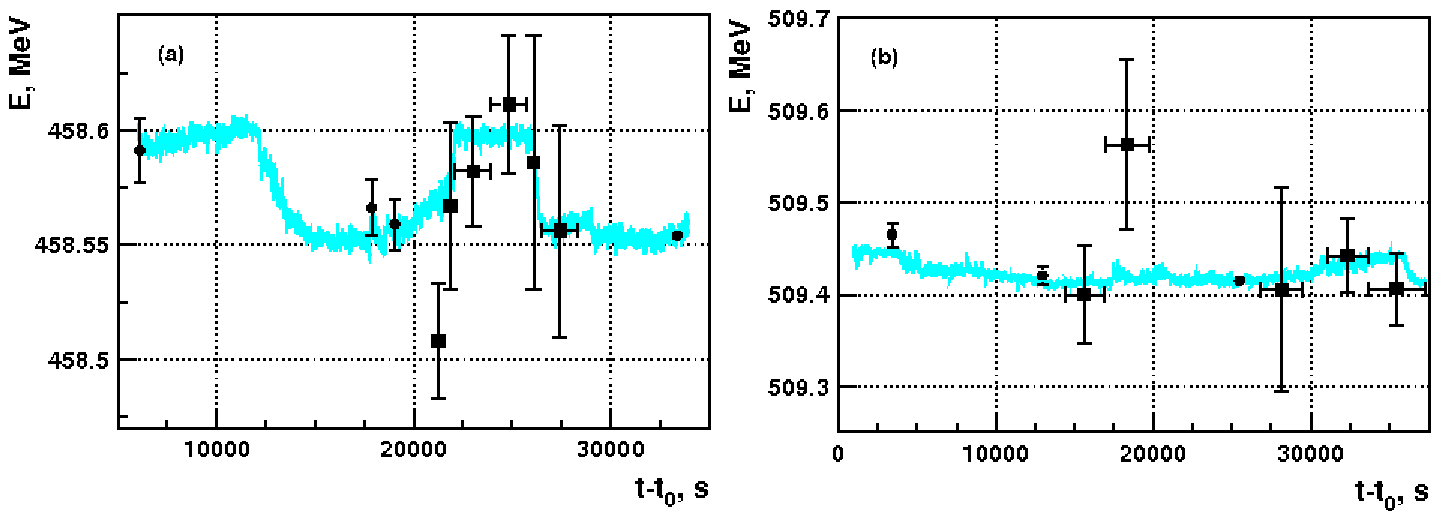}
\caption{Comparison of RD ($\bullet$) and CBS ($\blacksquare$) measurements.
The line shows the energy calculated using the magnetic field of the
collider.}
\label{rdpcbs}
\end{center}
\end{figure}

Four RD measurements were done at the beam energy 458 MeV 
(Fig.\ref{rdpcbs}(a)). Between RD measurements the electron beam was injected
in VEPP-2000 and its energy was measured using the CBS method. The collider
energy between RD calibrations was controlled by measuring fields in the
bending magnets by NMR sensors. Currents in the dipole correctors were 
also taken into account. Using these data the collider energy was calculated:
\begin{equation}
E_{NMR}=\alpha_c \bar{H},
\end{equation}
where $\bar{H}$ is the field in the VEPP-2000 ring, 
$\alpha_c=E_{RDP}/\bar{H}_{RDP}$, $E_{RDP}$ is the energy obtained by RD,
$\bar{H}_{RDP}$ is the field value at the moment of depolarization. The
$\alpha_c$ values for different measurements are in statistical agreement.
After averaging the relative statistical error of $\alpha_c$ is about
$4\times 10^{-6}$. The relative systematical error of $10^{-5}$ was estimated
taking into account unstability of the guiding magnetic field and presence of 
longitudinal magnetic fields. The dependence of energy on time 
(Fig.\ref{rdpcbs}) is due to a change of the dipole magnet temperature during 
operation of the complex.

The accuracy of the CBS measurement is estimated as 
\begin{equation}
\Delta E_{CBS}=E_{NMR}-E_{CBS}.
\end{equation}
The values of $\Delta E_{CBS}$ for various measurements are in agreement and
after averaging
\begin{equation}
\Delta  E_{CBS}= 0.023\pm 0.013 \mbox{~~MeV}.
\end{equation}
Taking into account this deviation, the relative accuracy of the beam energy
determination by the CBS method can be estimated as
$\Delta E_{CBS}/E_{CBS}\simeq 6\times 10^{-5}$. Similar energy measurements
at 509 MeV (Fig.\ref{rdpcbs}(b)) agree with this estimation.

\section{Conclusion}

The energy measurement system of the VEPP-2000 collider beam based on
the Compton backscattering method was designed, constructed, and 
commissioned. The systematic error of the beam energy
determination is tested by comparison with a measurement using
the resonance depolarization method and is estimated as $6\cdot 10^{-5}$.

\section*{Acknowledgment}
The authors are grateful to V.E. Blinov, S.I. Eidelman, A.G. Kharlamov, 
B.I. Khazin, S.I. Serednyakov, E.P. Solodov, Yu.A. Tikhonov, Yu.V. Usov
for supporting the
work and to P.L. Chapovsky, E.A. Sherbitsky and G.V. Tuliglovich for help in 
precise calibration of the laser wavelength. The work was supported by the
Ministry of Education and Science of the Russian Federation, by the RF
Presidential Grant for Scientific Schools NSh-6943.5320.2012.2 and by the RFBR
Grants No 13-02-00418-a, No 11-02-00276-a.


\begin{thebibliography}{99}
\bibitem{vepp2000}
D.E. Berkaev et al., Zh. Eksp. Teor. Fiz. 140 (2011) 247
\bibitem{rb}
V.V. Danilov et al., in Proc. EPAC'96, Barcelona, 1996, p.1149
\bibitem{okp1}
T. Yamazaki et al., IEEE Trans. on Nucl. Sci., Vol. NS-32, No5, 1985, p.3406
\bibitem{okp2}
Ian C. Hsu et. al., Nucl. Instr and Meth. A 384 (1997) 307; \\
Phys. Rev. E 54 (1996) 5657
\bibitem{bessy1}
R. Klein et al., Nucl. Instr. Meth. A 384 (1997) 293; \\
J. Synchrotron Rad. 5 (1998) 392
\bibitem{bessy2}
R. Klein et al., Nucl. Instr. Meth. A 486 (2002) 545
\bibitem{vepp4}
O.V. Anchugov et al., Zh. Eksp.Teor. Fiz. 136 (2009) 690 
[J.Exp.Theor.Phys. 109 (2009) 590]
\bibitem{bems}
E.V. Abakumova et al., Nucl. Instr. Meth. A 659 (2011) 21
\bibitem{rd}
A.N. Skrinsky and Yu.M. Shatunov, Sov. Phys. Uspekhi 32 (1989) 548
\bibitem{CN}
O. Klein and T. Nishina, Z. Phys. 52 (1929) 853
\bibitem{PRL}
E.V. Abakumova et al., Phys. Rev. Lett. 110 (2013) 140402
\bibitem{vac}
E.V. Abakumova et al., in Proc. of 18th Vacuum Congress, Phys. Proc. 32
(2012) 753
\bibitem{vp}
E.V. Abakumova et al., Vacuum Technic and Technology 20(2) (2010) 77 (in
Russian)
\bibitem{pdg}
J. Beringer et. al. (Particle Data Group), Phys. Rev. D86 (2012) 010001 
\bibitem{baier}
V.N. Baier, Usp. Fiz. Nauk 105 (1971) 3
\bibitem{g-2e}
I.B. Vasserman et al., Phys. Lett. B198 (1987) 302 
\end{thebibliography}
\end{document}